\documentclass[aps, prb, onecolumn, amsmath, amssymb, 10pt]{revtex4-2}

\usepackage{graphicx}
\usepackage{dcolumn}
\usepackage{bm}
\usepackage{color}
\usepackage[T1]{fontenc}
\usepackage[utf8]{inputenc}

\begin{document}

\title[]{Bound state in the continuum in an anisotropic photonic crystal supported by full-wave phase plate}

\author{Pavel S. Pankin$^{\dagger, *}$, Dmitrii N. Maksimov$^\dagger$, Ivan V. Timofeev}

\affiliation{
Kirensky Institute of Physics, FRC KSC SB RAS,Krasnoyarsk, 660036, Russia \\
Siberian Federal University, Krasnoyarsk, 660041, Russia\\
{$^\dagger$}These authors contributed equally\\
$^*$pavel-s-pankin@iph.krasn.ru}


\date{\today}

\begin{abstract}
We consider bound states in the continuum (BICs) in a 1D multilayered system of an anisotropic defect layer embedded into an anisotropic photonic crystal. We analytically demonstrate that an anisotropic defect layer embedded into anisotropic photonic crystal supports accidental BICs. These BICs can be transformed to a high-Q resonances by variation of one of the system's parameters. At the same time the BICs are remarkably robust in a sense that a true BIC can be recovered by further tuning any of the other system's parameters leading to tunability of the resonance position.
\end{abstract}

\maketitle

\section{Introduction}

Multilayered microcavities based on 1D photonic crystals (PhCs) have found a broad range of applications in modern photonics. The microcavities are widely used in lasers, thermal emitters, single photon sources, light emitting diodes, filters, sensors, solar cells, and absorbers~\cite{Joannopoulos2008bk, kavokin2017microcavities}. The problem of external control of multilayered microcavity resonances addresses their spectral positions and quality factors (Q-factors). The first part of the problem is solved by adjustment of the optical thickness of the layers that shifts the spectral position. Various functional materials have been proposed for this purpose~\cite{Chang2012TempPC, Yoon2006DyeDefPC, Aly2020GraphenePC, Aly2017FaradayPC, Chaves2021PressurePC, Goto2008} specifically liquid crystals \cite{Vetrov2001MC, Ozaki2003LCdef, Arkhipkin2008, Arkhipkin2011, Pankin2021APL}. For the second part, the control of the Q-factor can be achieved by utilizing recently emerged bound states in the continuum (BICs) \cite{HsuChiaWei2016, Koshelev2019Review, Azzam2020BICReview, sadreev2021interference, Joseph2021BICReview}.

Application of BICs allows for engineering optical modes with tunable Q-factor through controlling the energy decay rates from the localized state into the continuum of scattering channels by varying parameters of the BIC host structure. The Q-factor can reach extremely high values on approach to the BIC being only restricted by material losses and fabrication inaccuracies. By now several models for controlling the Q-factor in multilayered structures have been proposed theoretically~\cite{Timofeev2018_BIC, Ignatyeva2020BIC, Pankin2020Fano} and implemented experimentally~\cite{GomisBresco_Torner2017_BIC1D, Pankin2020BIC}.

In this paper we consider BICs of interference nature (accidental~\cite{Hsu2013} or Friedrich-Wintgen\cite{Friedrich1985} BIC mechanisms) in a 1D multilayered system. The system under scrutiny was proposed in our previous papers where we found that an anisotropic defect layer (ADL) embedded into an anisotropic PhC can support symmetry protected BICs \cite{Timofeev2018_BIC, Pankin2020Fano}. Later a similar system was implemented experimentally with an ADL sandwiched between metallic mirror and a PhC arm operating at Brewster's angle ~\cite{Pankin2020BIC}. Here, we further extend our studies by analytically demonstrating that the PhC embedded ADL supports accidental BICs which are robust to variation of optical and geometric parameters. 

\section{The System}
\begin{figure}[ht]
\center{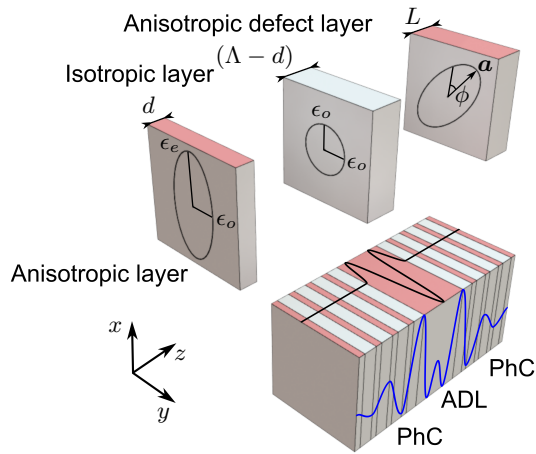}
\caption{
{One-dimensional PhC structure stacked of alternating layers
of an isotropic dielectric material with permittivity $\epsilon_o$
(gray) and an anisotropic material with the permittivity
components $\epsilon_o$ and $\epsilon_e$ (pink). An anisotropic
defect layer with a tuneable permittivity tensor is inserted in
the center of the structure. The analytic solution for the
BIC mode profile 
is plotted on top and right sides of the stack: the $x$-wave component
Re$(E_x)$ -- blue, the $y$-wave component Re$(E_y)$ -- black.}}
\label{fig1}
\end{figure}

We consider the system of an ADL of thickness $L$ embedded into an anisotropic PhC
as shown in Fig.~\ref{fig1}. Each PhC
arm is a one-dimensional PhC with alternating layers of isotropic
and anisotropic dielectrics. The layers are arranged along
the $z$-axis with period $\Lambda$. The isotropic layers are made
of a dielectric material with permittivity $\epsilon_o$ and
thickness $\Lambda-d$. The thickness of each anisotropic layer is
$d$. The principal dielectric axes of the anisotropic layers are
aligned with the $x$, $y$-axes. The corresponding permittivity
component constants are denoted by $\epsilon_{e}$,
$\epsilon_{o}$. The ADL is made of the same anisotropic material. The principal axes of the ADL are tilted with
respect to the principal axes of the PhC arms as shown in Fig.
\ref{fig1}. Propagation of the monochromatic electromagnetic waves at normal incidence is
governed by Maxwell's equations of the following form \cite{Landau8_1984}

\begin{equation}
\left\{\begin{array}{cc}
0 & \nabla \times \\
-\nabla \times & 0
\end{array}\right\}
\left\{\begin{array}{c}
{\bm{E}} \\
{\bm{H}}
\end{array} \right\}=
-ik_0
\left\{\begin{array}{c}
\hat{\epsilon} {\bm{E}} \\
{\bm{H}} \end{array}\right\},
\label{Maxwell}
\end{equation}
where $\bm{E}$ is the electric vector, $\bm{H}$ is the magnetic vector, $k_0 = \omega/c$ is the wave number in vacuum
with $c$ as the speed of light, and, finally, $\hat{\epsilon}$ is the dielectric tensor.
The orientation of the ADL optical axis is
determined by the unit vector
\begin{equation}
\bm{a} = [\cos{(\phi)}, \sin{(\phi)}, 0],
\label{a}
\end{equation}
as shown in Fig. \ref{fig1}.
Since the reference frame is aligned with the
optical axes in the PhC, the dielectric tensor is diagonal everywhere out of the ADL.
Given that $\bm{a}$ is specified by the tilt angle $\phi$, in the ADL it takes the following
form
\begin{equation}
\hat{\epsilon} = \left\{\begin{array} {cc}
\epsilon_e \cos^2 (\phi) + \epsilon_o \sin^2 (\phi) & \sin{(2\phi)} \;  (\epsilon_e-\epsilon_o)/2 \\
\sin{(2\phi)}  \; (\epsilon_e-\epsilon_o)/2 & \epsilon_e \sin^2 (\phi) + \epsilon_o \cos^2 (\phi)
\end{array} \right\}.
\label{eq:epsilon}
\end{equation}

\begin{figure}[ht]
\centering\includegraphics{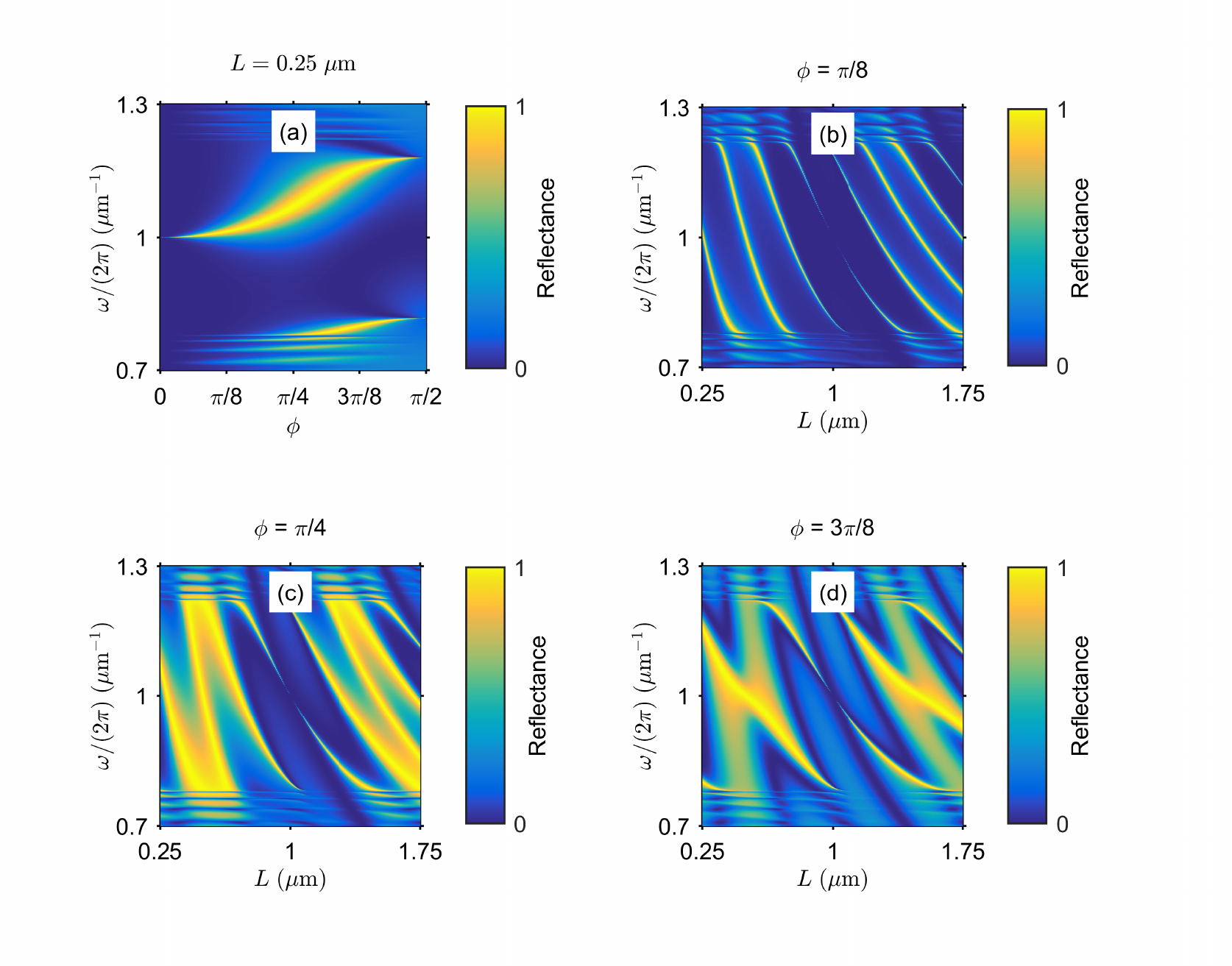}
\caption{Reflectance spectra computed by Berreman transfer matrix method for the structure containing 20 periods in each PhC arm. (a) Reflectance against the incident frequency and the ADL rotation angle $\phi$ at constant ADL thickness $L = 0.250 \mu$m.  Collapsing Fano resonance at  $\phi = 0, \pi/2$ corresponds to  symmetry-protected BICs. (b,c,d) Reflectance against the incident frequency and the ADL thickness $L$ at fixed rotation angle $\phi = \pi/8$ (a), $\pi/4$ (b), $3\pi/8$ (d). Collapsing resonance at $L = 1 \mu$m corresponds to accidental BIC. The parameters are $\epsilon_e
= 4$, $\epsilon_o = 1$, $d = 0.125 \ \mu m$, $(\Lambda - d) =
0.250 \ \mu m$.}
\label{fig2}
\end{figure}

The dispersion of waves in the PhC arms depends on polarization.
For the $x$-polarized
waves ($x$-waves) the dispersion relationship is identical to that of a one-dimensional PhC \cite{Rytov1956,
YarivYeh1984bk, ShiTsai1984_PBG, CamleyMills1984_PBG}
\begin{equation}
\cos{(K \Lambda)} = \cos{(k_{e} d)} \cos{[k_{o} (\Lambda - d)]} - \frac{1 + r_{o e}^2}{1 - r_{o e}^2} \sin{(k_{e} d)} \sin{[k_{o}
(\Lambda - d)]},
\label{Rytov}
\end{equation}
where $K$ is the Bloch wave number,
\begin{equation}
k_{e} = k_0 \sqrt{\epsilon_e} = k_0 n_e, \ k_{o} = k_0 \sqrt{\epsilon_o} = k_0 n_o,
\label{o_and_e}
\end{equation}
and the Fresnel coefficient $r_{o e}$ is given by
\begin{equation}
r_{o e} = \frac{k_{o} -  k_{e}} {k_{o} + k_{e}}.
\label{Fresnel_ab_perp}
\end{equation}
Equation (\ref{Rytov}) defines the band structure for the
$x$-waves. 
In Fig. \ref{fig2}(a) we
demonstrate the reflectance spectrum of the system with $20$
bi-layers in each PhC arm; the overall system being submersed in
air. PhC arms are always transparent to the $y$-polarized waves ($y$-waves) with dispersion
$k_o=\epsilon_o k_0$. Notice, though, that the $y$-waves
transmittance exhibits a dip at the band gap. This
dip is due to a high quality resonant mode predicted in
\cite{Timofeev2018_BIC}. One  can see that at $\phi = 0, \pi/2$, the resonant line collapses indicating a symmetry protected BIC. As it can be seen from the spectra in Fig.~\ref{fig2}~(b-d) the increase of the ADL thickness also leads to collapsing Fano features at ADL rotation angles different from $\phi = 0, \pi/2$. Thus, we conclude that the BIC persists despite the symmetry is broken, i.e. the BIC occurs via the accidental or Friedrich-Wintgen BIC mechanisms. It is worth to point out that under given $n_o$ and $n_e$ the ADL thickness leading to the Fano feature collapse, does not depend on the ADL rotation angle $\phi$.

\section{Resonant eigenmode}

The resonant states are the eigenmodes of Maxwell's equations (\ref{Maxwell}) with reflectionless boundary conditions
in the PhC arms. The equation for resonant eigenfrequencies can be obtained by matching the general solution in the ADL
to all outgoing waves in the PhC arms.


The unit vector along the propagation direction is defined as
\begin{equation}
\bm{\kappa}^{\scriptscriptstyle(\pm)}  = [0, 0, \pm 1],
\label{kappa_eo}
\end{equation}
where the symbol $\pm$ designates forward and backward propagating waves with respect to
the $z$-axis. The ADL supports two types of electromagnetic
waves of different polarization. The $e$-waves with wavenumber $k_e=\epsilon_e k_0$ are polarized along the director $\bm{a}$,
equation (\ref{a}), while for the
$o$-waves the wavenumber $k_o=\epsilon_o k_0$ is perpendicular to both $\bm{a}$ and $\bm{\kappa}$.
The electric and magnetic vectors of the $e$-wave can be written as
\begin{equation}
\bm{E}_{e}^{\scriptscriptstyle(\pm)} = E_{e}^{\scriptscriptstyle(\pm)} \bm{a}, \ \ \bm{H}_{e}^{\scriptscriptstyle(\pm)}
= \frac{k_e}{k_0}\left[\bm{\kappa}^{\scriptscriptstyle(\pm)} \times \bm{E}_{e}^{\scriptscriptstyle(\pm)} \right],
\label{e_EH}
\end{equation}
where $E_{e}^{(\pm)}$ are unknown amplitudes.
At the same time for $o$-waves we have
\begin{equation}
\bm{E}_{o}^{\scriptscriptstyle(\pm)} = E_{o}^{(\pm)} \left[ \bm{a} \times \bm{\kappa}^{\scriptscriptstyle(\pm)} \right], \ \
\bm{H}_{o}^{\scriptscriptstyle(\pm)} = \frac{k_o}{k_0}\left[\bm{\kappa}^{\scriptscriptstyle(\pm)} \times \bm{E}_{o}^{\scriptscriptstyle(\pm)}
 \right],
\label{o_EH}
\end{equation}
where $E_{o}^{(\pm)}$ are again unknown amplitudes.
The general solution of equations (\ref{Maxwell}) in the ADL, $\ z \in [-L/2,\ L/2]$, is written as a sum of
the forward and backward propagating waves
\begin{equation}
\bm{E} = \sum_{j=o,e} \left(\bm{E}_{j}^{\scriptscriptstyle(+)} e^{i k_{j} z} + \bm{E}_{j}^{\scriptscriptstyle(-)} e^{-i k_{j} z} \right),\ \
\bm{H} = \sum_{j=o,e} \left(\bm{H}_{j}^{\scriptscriptstyle(+)} e^{i k_{j} z} + \bm{H}_{j}^{\scriptscriptstyle(-)} e^{-i k_{j} z} \right).
\label{sum_EH}
\end{equation}

The general solution of Maxwell's equations (\ref{Maxwell}) in the PhC arms is also written as a sum of forward
and backward  propagating waves. For
the $x$-waves the field components $E_x$ and $H_y$ in the isotropic layer
with the cell number $m$, $\ z \in [L/2 + m \Lambda,\ L/2 + (m + 1) \Lambda - d]$, are written as
\begin{equation}
\begin{aligned}
& E_{x}^{(m)} =  e^{iK \Lambda m}\left[A^{\scriptscriptstyle(+)} e^{i k_{o} (z - L/2 - m  \Lambda)}
+ A^{\scriptscriptstyle (-)} e^{-i k_{o} (z - L/2 - m \Lambda)}\right], \\
& H_{y}^{(m)} = \frac{k_{o}}{k_0} e^{iK \Lambda m}\left[A^{(+)} e^{i k_{o}
 (z - L/2 - m \Lambda)} - A^{(-)} e^{-i k_{o} (z - L/2 - m \Lambda)}\right].
\end{aligned}
\label{xEH1}
\end{equation}
In the anisotropic layer with the cell number $m$, $\ z \in [L/2 + (m + 1) \Lambda - d,\ L/2 + (m + 1) \Lambda] $, we have
\begin{equation}
\begin{aligned}
& E_{x}^{(m)}  =  e^{iK \Lambda m}\left[B^{(+)} e^{i k_{e} (z - L/2 - (m + 1) \Lambda + d)} + B^{(-)} e^{-i k_{e} (z -  L/2 - (m + 1) \Lambda + d)}\right], \\
& H_{y}^{(m)} = \frac{k_{e}}{k_0} e^{iK \Lambda m}\left[B^{(+)} e^{i k_{e} (z - L/2 - (m + 1) \Lambda + d)} - B^{(-)} e^{-i k_{e} (z -  L/2 - (m + 1) \Lambda + d)}\right].
\end{aligned}
\label{xEH2}
\end{equation}
By applying the continuity condition for the tangential field components
the solutions (\ref{xEH1}) and (\ref{xEH2}) are matched
on the boundary between the anisotropic layer in the $(m-1)_{\rm th}$ cell and the isotropic
layer in $m_{\rm th}$ cell,  $\ z = L/2 + m \Lambda$, as well as on the boundary between the layers
in the $m_{\rm th}$ cell, $\ z = L/2 + (m + 1)\Lambda - d$. This gives us a system of four equations for four
unknowns, $A^{(+)}, A^{(-)}, B^{(+)}, B^{(-)}$. After solving for $B^{(+)}$ and $B^{(-)}$, this system can
be reduced to the following two equations
\begin{equation}
\left\{\begin{aligned}
A^{(+)} \left[e^{i k_{o} (\Lambda - d)} - e^{iK \Lambda} e^{-i k_{e} d}\right] -
 A^{(-)} r_{o e}\left[e^{-i k_{o} (\Lambda - d)} - e^{iK \Lambda} e^{-i k_{e} d}\right] = 0, \\
-A^{(+)} r_{o e} \left[e^{i k_{o} (\Lambda - d)} - e^{iK \Lambda} e^{i k_{e} d}\right] +
A^{(-)} \left[e^{-i k_{o} (\Lambda - d)} - e^{iK \Lambda} e^{i k_{e} d}\right] = 0,
\end{aligned}\right.
\label{PC_equations}
\end{equation}
where $r_{oe}$ is given by equation (\ref{Fresnel_ab_perp}). One can easily check that Eq. (\ref{PC_equations}) is only solvable
when $K$ satisfies the dispersion relationship (\ref{Rytov}).

In contrast to the $x$-waves, for the outgoing $y$-waves in the right PhC arms the solution is given by
\begin{equation}
\begin{aligned}
& E_{y}  =
-C^{(+)} e^{i k_{o} (z - L/2)}, \\
& H_{x}  = \frac{k_{o}}{k_0} C^{(+)} e^{i k_{o} (z - L/2)}.
\end{aligned}
\label{yEH}
\end{equation}
Notice that so far we have not written down the solution in the left PhC arm. The direct application of that solution can be avoided
by using the mirror symmetry of the system.
In the antisymmetric case we have
\begin{equation}
\bm{E}(z) = - \bm{E}(-z),
\label{atysymmetry}
\end{equation}
while in the symmetric case the condition is:
\begin{equation}
\bm{E}(z) = \bm{E}(-z).
\label{symmetry}
\end{equation}

Now by matching equation
(\ref{sum_EH}) to both equation (\ref{xEH1}) and equation (\ref{yEH}) on the interface between the ADL and the
right PhC arm and using equations (\ref{PC_equations}, \ref{atysymmetry},\ref{symmetry}) one obtains eight equations for
eight unknown variables $E_{e}^{(+)}, E_{e}^{(-)}, E_{o}^{(+)}, E_{o}^{(-)}, A^{(+)}, A^{(-)}, C^{(+)}, K$.
After some lengthy and tedious calculations one finds that the system is solvable under the following condition
\begin{equation}
\frac{\xi e^{i k_{o} (\Lambda - d)} - r_{o e} e^{-i k_{o} (\Lambda - d)}}{\xi e^{-i k_{e} d} - r_{o e} e^{-i k_{e} d}} -
\frac{e^{-i k_{o} (\Lambda - d)} - \xi r_{o e} e^{i k_{o} (\Lambda - d)}}{e^{i k_{e} d} - \xi r_{o e} e^{i k_{e} d}} = 0,
\label{disp}
\end{equation}
where
\begin{equation}
\xi = -e^{i k_{o} L}\sin^2{(\phi)} + \frac{r_{oe} - e^{i k_{e} L}}{1 - r_{oe} e^{i k_{e} L}} \cos^2{(\phi)}
\label{x}
\end{equation} 
for the antisymmetric case and
\begin{equation}
\xi = e^{i k_{o} L}\sin^2{(\phi)} + \frac{r_{oe} + e^{i k_{e} L}}{1 + r_{oe} e^{i k_{e} L}} \cos^2{(\phi)}
\label{x_sym}
\end{equation}
for the symmetric case. Eqs. (\ref{disp}, \ref{x}, \ref{x_sym}) are solved for the complex eigenfrequency
\begin{equation}
\omega_r = \tilde{\omega}-i \gamma.
\label{eig}
\end{equation}

\begin{figure}[ht]
\centering\includegraphics{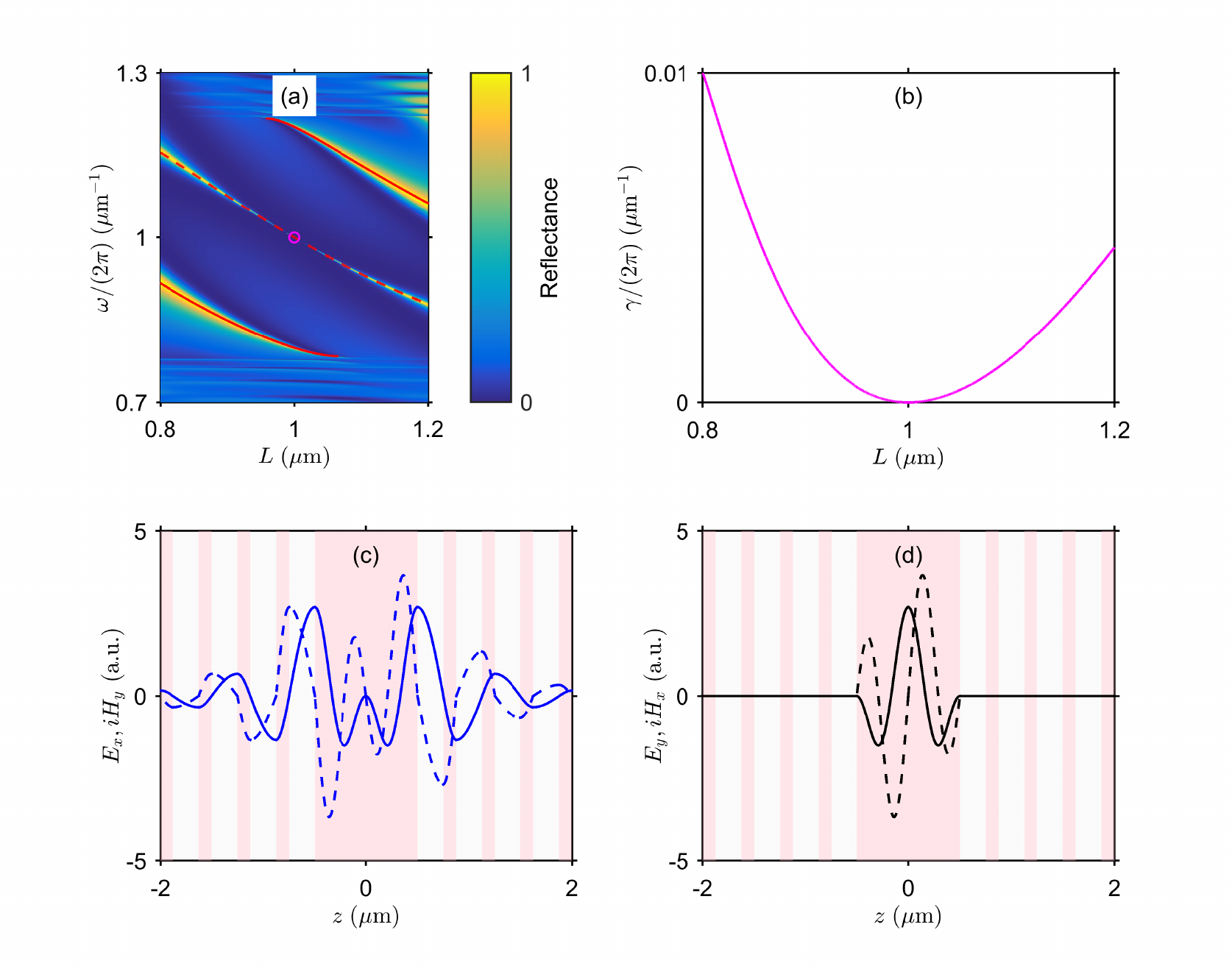}
\caption{(a) Zoomed in reflectance spectrum from Fig.~\ref{fig2}~(c). Solid red lines show the real part of solution of Eqs.~(\ref{disp}, \ref{x}) in the antisymmetric case. The dashed line demonstrates the symmetric solution obtained from Eqs.~(\ref{disp}, \ref{x_sym}). The magenta circle shows the position of the BIC from Eq.~(\ref{FWPP}). (b) The imaginary part of solution~(\ref{disp}, \ref{x_sym}). (c,d) BIC field profile (\ref{EHADL}-\ref{EHPhC_aniso}). All parameters are identical to that in the caption to Fig.~\ref{fig2}~(c).The normalization condition is used (\ref{Energy}) to yield $A = \sqrt{2\pi/7}$.} 
\label{fig3}
\end{figure}

The real part of the eigenfrequency $\tilde{\omega} $determines the position of the resonance. The position of the resonance is shown in Fig.~\ref{fig3}~(a) superposed with the reflectance spectrum. The imaginary part of the eigenfrequency which determines the width of the resonance $\Delta\omega = 2\gamma$ predicts a collapsing Fano feature in the same point as found from the numerical reflectance spectrum Fig.~\ref{fig3}~(b). Let us show that the point of collapse corresponds to a BIC.

\section{Bound state in the continuum}

For finding solution for accidental BIC in the symmetric case we match Eqs. (\ref{sum_EH}, \ref{xEH1}, \ref{yEH}) on the boundary between the ADL and the PhC arms using (\ref{PC_equations}) and (\ref{symmetry}) together with the condition on destructive interference between ordinary and extraordinary waves projected on the $y$ axis

\begin{equation}
E_{y}(L/2) = 0, \ H_{x}(L/2) = 0.
\label{BIC_condition}
\end{equation}

The solution for the BIC takes the following form

\begin{equation}
n_o\tan{(k_o L/2)} - n_e\tan{(k_e L/2)} = 0.
\label{BIC_solution}
\end{equation}

 Equation~(\ref{BIC_solution}) transforms to identity when two conditions below are simultaneously fulfilled
\begin{equation}
k_o L/2 = m_1 \pi, \ k_e L/2 = m_2 \pi, \ m_{1,2} \in \mathbb{N}.
\label{BIC_solution2}
\end{equation}
In the simplest case $m_2 = m_1 + 1$ we have a full wave phase plate  condition
\begin{equation}
(k_e - k_o) L = 2 \pi.
\label{FWPP}
\end{equation}
The solution~(\ref{FWPP}) shown in Fig.~\ref{fig3}(a) by a magenta circle corresponds to collapse of the Fano feature. The full wave phase plate preserves the linear polarization of the incident wave independent of orientation of the optical axis. Therefore the tilt angle
is absent from both final solution for the BIC, Eq. (\ref{BIC_solution}), and the reflectance spectra in Fig.~\ref{fig2}.

\begin{figure}[ht]
\centering\includegraphics{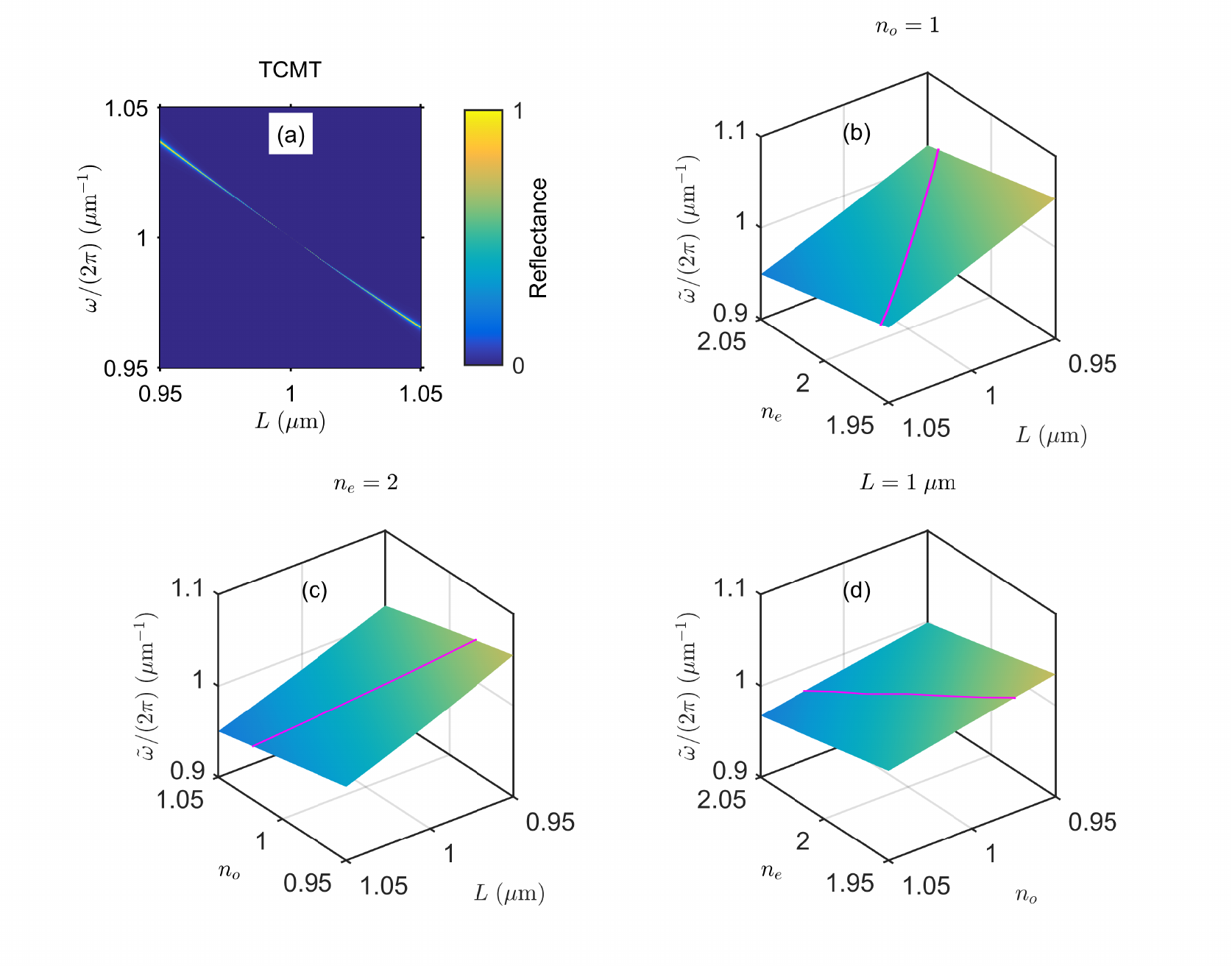}
\caption{(a) Reflectance spectrum obtained by the TCMT, Eq.~(\ref{TCMT}). (b-d) The real part of solution of Eqs.~(\ref{disp}, \ref{x_sym}) in the symmetric case. Magenta lines show the positions of the BIC ($\gamma = 0$). The parameters are the same as in the caption to Fig.~\ref{fig2}~(c).}
\label{fig4}
\end{figure}

The BIC field profile inside the ADL
\begin{equation}
\left\{\begin{aligned}
& E_x = 2 A (\cos{(k_e z)} \cot{(\phi)} \cos{(\phi)} - \cos{(k_o z)} \sin{(\phi)}), \\
& H_y = 2i A (n_e \sin{(k_e z)} \cot{(\phi)} \cos{(\phi)} - n_o \sin{(k_o z)} \sin{(\phi)}), \\
& E_y = 2 A (\cos{(k_e z)} + \cos{(k_o z)})  \cos{(\phi)}, \\
& H_x = - 2i A (n_e \sin{(k_e z)} + n_o \sin{(k_o z)}) \cos{(\phi)}.
\label{EHADL}
\end{aligned}\right.
\end{equation}
Inside the PhC arms, in the isotropic layer
with the cell number $m$, $\ z \in [L/2 + m \Lambda,\ L/2 + (m + 1) \Lambda - d]$ we have
\begin{equation}
\left\{\begin{aligned}
& E_x^{(m)} = \frac{2A}{\sin{(\phi)}} (-q)^m \cos{(k_o (z - L/2 - m \Lambda))}, \\
& H_y^{(m)} = \frac{2iA}{\sin{(\phi)}} (-q)^m n_o \sin{(k_o (z - L/2 - m \Lambda))}, \\
& E_y = 0, \\
& H_x = 0.
\label{EHPhC_iso}
\end{aligned}\right.
\end{equation}
Finally, inside the PhC arms, in the anisotropic layer with the cell number $m$, $\ z \in [L/2 + (m + 1) \Lambda - d,\ L/2 + (m + 1) \Lambda] $, we have
\begin{equation}
\left\{\begin{aligned}
& E_x^{(m)} = \frac{2A}{\sin{(\phi)}} (-q)^{m+1} \sin{(k_e (z - L/2 - (m + 1) \Lambda + d))}, \\
& H_y^{(m)} = -\frac{2iA}{\sin{(\phi)}} (-q)^{m+1} n_e \cos{(k_e (z - L/2 - (m + 1) \Lambda + d))}, \\
& E_y = 0, \\
& H_x = 0.
\label{EHPhC_aniso}
\end{aligned}\right.
\end{equation}

Here $q = n_o/n_e$, the amplitude $A$ has to be defined from a proper normalization condition, for example by equating the total energy of BIC ${\cal E}$ to unity:

\begin{equation}
{\cal E}=\frac{1}{8\pi}\int\limits_{-\infty}^{+\infty}dz \left[
\left.{E}\right.^{\dagger}\hat{\epsilon}(z){E}+
\left.{H}\right.^{\dagger}{H} \right] = 1,
\label{Energy}
\end{equation}
where  
\begin{equation}
\bm{E} = \left\{\begin{array}{c}
E_x \\
E_y
\end{array} \right\},\ \bm{H} = \left\{\begin{array}{c}
H_x \\
H_y
\end{array} \right\}.
\label{EH}
\end{equation}
The BICs field profile is shown in Figs.~\ref{fig1} and \ref{fig3}~(c,d).

\section{TCMT and robustness of the BIC}

The general expression for the reflection/transmission amplitudes was
obtained in \cite{Timofeev2018_BIC} based on temporal coupled mode theory \cite{FanShanhui2003} as
\begin{equation}
\rho = \frac{i\gamma}{(\tilde{\omega}-\omega) +i\gamma}, \ \tau=1-\rho,
\label{TCMT}
\end{equation}
where $\tilde{\omega}$ and $\gamma$ are real and imaginary parts of the resonant eigenfrequency~(\ref{eig}), and $\omega$ is the incident frequency.
The single resonance solution (\ref{TCMT}) coincides with the numerical data to a good accuracy, see Fig.~\ref{fig4}~(a).

According to Eq. (\ref{TCMT}) the reflectance depends only on two parameters $\gamma$ and $\tilde{\omega}$. The radiative decay rate $\gamma$ can be controlled by detuning one of the system's parameters from the BIC point, see Fig. \ref{fig2}. Technically both $\tilde{\omega}$ and $\gamma$ can be found from solving Eqs.~(\ref{disp}, \ref{x_sym}). In Figs. \ref{fig4}~(b,c,d) we show the position of the BIC as function any pair of two parameters $L, n_e, n_o$. One can see that the positions of the BIC form continuous lines in all the considered cases. In other words after variation of a single parameter any other parameter can be tuned to recover a BIC making it robust with respect to parameter variation. 

\section{Conclusion}

In this paper we considered BICs in a 1D multilayered system of an anisotropic defect layer embedded into an anisotropic PhC. We analytically demonstrated that the PhC-embedded full-wave phase plate ADL supports accidental BICs. These BICs can be transformed to a high-Q resonances by variation of one of the system's parameters allowing for control of the resonant width. At the same time the BICs are remarkably robust in a sense that a true BIC can be recovered by further tuning any of the other system's parameters leading to tunability of the resonance position. The demonstrated effect will lead to robustness of the BIC supporting set-up with respect to fabrication tolerance paving a way for implementing microcavities with tunable Q-factor.

\section*{Acknowledgments}

The authors are grateful to V.A.~Stepanenko for helpful discussions. P. S. Pankin is grateful for the support of the President of the Russian Federation grant MK-4012.2021.1.2.

\bibliography{FW_BIC.bib}

\end{document}